# Improving Community Resiliency and Emergency Response With Artificial Intelligence

**Ben Ortiz**
Accenture Federal Services[*]

**Laura Kahn**[†]
Accenture Federal Services
laura.kahn@accenturefederal.com

**Marc Bosch**
Accenture Federal Services

**Philip Bogden**
Accenture Federal Services

**Viveca Pavon-Harr**
Accenture Federal Services

**Onur Savas**
Accenture Federal Services

**Ian McCulloh**
Accenture Federal Services

**ABSTRACT**

New crisis response and management approaches that incorporate the latest information technologies are essential in all phases of emergency preparedness and response, including the planning, response, recovery, and assessment phases. Accurate and timely information is as crucial as is rapid and coherent coordination among the responding organizations. We are working towards a multi-pronged emergency response tool that provide stakeholders timely access to comprehensive, relevant, and reliable information. The faster emergency personnel are able to analyze, disseminate and act on key information, the more effective and timelier their response will be and the greater the benefit to affected populations. Our tool consists of encoding multiple layers of open source geospatial data including flood risk location, road network strength, inundation maps that proxy inland flooding and computer vision semantic segmentation for estimating flooded areas and damaged infrastructure. These data layers are combined and used as input data for machine learning algorithms such as finding the best evacuation routes before, during and after an emergency or providing a list of available lodging for first responders in an impacted area for first. Even though our system could be used in a number of use cases where people are forced from one location to another, we demonstrate the feasibility of our system for the use case of Hurricane Florence in Lumberton, a town of 21,000 inhabitants that is 79 miles northwest of Wilmington, North Carolina.

**INTRODUCTION**

There is a need for a multi-pronged crisis planning and response system that incorporates several ways to layer, encode and visualize relevant information and improve human decision-making (**Van de Walle**, 2007). We propose a multi-pronged artificial intelligence (AI) emergency tool to improve a community's resilience to natural disasters such as hurricanes, wild fires, earthquakes and other types of emergencies or crisis events. As the magnitude and impact of emergencies continues to increase, having tools to speed human decision-making is critical. In the age of big data and AI, emergency management personnel can benefit from using multiple data sources and visualizations to improve their decisions in a crisis context. AI provides mechanisms to process this large amount of data at speed which gives users reliable information more quickly, which in turn can accelerate emergency planning and response. Our tool improves a community's resilience in times of crisis. Community resilience is defined as "the ability to adapt through the redevelopment of the community in ways that reflect the community's values, and goals, and its evolving understanding of external forces with which it must contend". By encoding input data from a variety of sources, including post-event damages from automated overhead images, an accurate assessment of infrastructure damages can be made. All of these heterogeneous data layers work together for a more accurate risk assessment. In this work we have focused on building tools for hurricane disaster response though our approach could be re-purposed or expanded to other emergency events or use cases with minimal structural changes.

---

[*]https://www.accenture.com/us-en/services/us-federal-government/artificial-intelligence

[†]corresponding author

In particular, Hurricane Florence is the wettest tropical cyclone on record in the Carolinas and made landfall September 18, 2018 (**Florence**, 2018). According to the U.S. Geological Survey (USGS), most of the damage caused in Lumberton and other areas in the Carolinas during Hurricane Florence were due to inland freshwater flooding. The Lumber River runs through the city and crested twice at over 20 feet, well above the flood stage of 13 feet. The water flooded the city, highways and surrounding rural areas. More than 500 people were evacuated and more than 500 structures were damaged. Accounting for the uptick in more frequent inland flooding is a critical consideration in a modern, effective emergency ecosystem.

In this work we give an update of our current AI-driven emergency management tool with focus on adding and using additional data layers and techniques. Our tool uses automated image analysis and inland flood modeling to produce a system that augments human decision-making for faster and accurate evacuation routing before, during and after crisis events, and has a feature to populate a list of lodging options in the affected area given the current observed conditions. Though our work has not been deployed to real users in an emergency communication context and is not introducing new algorithms, we feel its scientific contribution is to add relevant information to maps prior to, during and after a natural disaster to augment human decision-making.

**RELATED WORK**

Inspired by previous work, we have built a tool that uses AI to encode several layers of data, has a user-centered design framework to drive technical requirements and uses a shortest route Dijkstra machine learning algorithm to improve route planning and communication before, during and after natural hazards. (**Hilljegerdes**, 2019) uses agent-based modeling to calculate the shortest walking distance of evacuees to shelters prior to a hurricane. (**Zachry**, 2015) describes a SLOSH model that provides probabilistic coastal storm surge model forecasts using a sampling of meteorological forecasts. The National Water Model (NWM) provides coastal flooding information but was not operational following Hurricane Florence in 2018 and does not model inland flooding (US Geological Survey Flood Inundation Map Program). We also leveraged previous efforts to provide a method of translating aerial image data into segmented and classified objects such as buildings and roads using semantic segmentation computer vision techniques. Similar techniques have also been proposed to exploit satellite imagery to find a variety of features such as buildings and roads (**Prasad**, 2016, **Wurm**, 2019, **Bosch**, 2019). None of the research to date combines inland inundation modeling with computer vision semantic segmentation data layers to be near real-time inputs for autonomous decision-making (such as route evacuation calculations with machine learning algorithms) in a crisis scenario.

**THE EMERGENCY RESPONSE TOOL**

As shown in figure 1, a crisis or emergency event causes more uncertainty and more frequent communications. Our multi-pronged emergency tool combines several techniques to reduce uncertainty, improve communication and yields actionable insights from the data. We believe that combining several components such as inundation level mapping and computer vision semantic segmentation and using them as inputs for calculating and displaying the optimal evacuation route is a novel approach to create actionable information from data. As shown in figure 2, the five heterogeneous input data sources include Digital Elevation Model (DEM) land topography data, semantic segmentation flood data, Open Street Map geospatial information (GIS) data, shelter location GIS data and RoadNet GIS data. The input data sources are encoded using R software and then road flooding information is provided to the user. The road flooding information shows which roads are flooded in a given area and form the hazard risk for road analysis. After the input data is encoded, the information from that data can help a user with a risk assessment which could include hazard risk for the roads, area and population. The risk assessment layer of our emergency system output a display of damaged and non-damaged roads, which can augment human decision-making and speed up decision-making time. Our tool increases a community's resilience to natural disasters and other types of emergencies, as well as improves emergency management personnel's response by giving them the right information at the right time. Our tool can therefore improve overall resiliency to emergency events by improving decision-making under uncertainty. In this section we describe our main components: inundation maps, semantic segmentation for flooding detection, and the AI-driven front-end interface for routing and lodging recommendation.

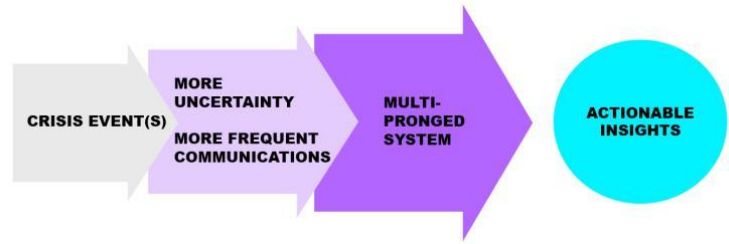

**Figure 1. Effects of Emergencies**

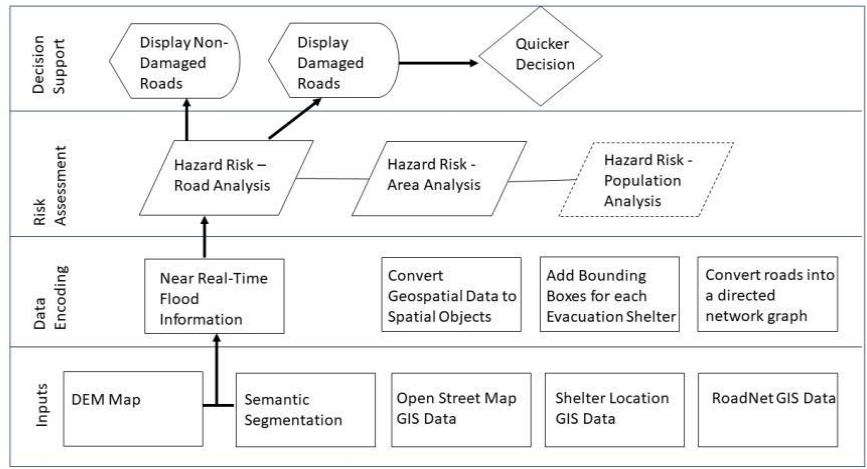

**Figure 2. Components of the Emergency System**

**Inundation Maps**

The first component of our holistic emergency resilience AI tool is a custom-made inundation map. According to the USGS, a flood inundation map "shows where flooding may occur over a range of water levels in the community's local stream or river" (**USGS Flood Inundation Map Program**, 2018). Existing models focus on coastal rather than inland flooding to show communities where flooding is likely to occur. Additionally, the National Oceanic and Atmospheric Administration (NOAA) National Hurricane Center provides output from the Sea, Lakes and Overland Surges from Hurricanes otherwise known as a SLOSH model (**Zachry**, 2015). The SLOSH model includes the effects of hurricane winds along the coast but it does not include the effects of precipitation on land from the prolonged extreme rains during a hurricane such as Hurricane Florence. The SLOSH model is inadequate to determine inland flooding so more complex NOAA National Water Models are the closest operational approximation data (**USGS Digital Elevation Models**, 2018). Version 1.2 (March 2018) of the NWM currently predicts stream flow for the United States, but operational inundation maps for inland flooding across the entire continental U.S. are still being tested. The NWM lacks post-disaster verified road flooding data. Since operational inundation maps were not available for Hurricane Florence in 2018, we used USGS Digital Elevation Models (DEMs) as a proxy. Even when Version 1.3 of the NWM becomes available, it will lack information on actual conditions of the roads and structures post-emergency (**Carswell**, 2005). Low lying areas in the DEM have spatial characteristics that determine the structure in inundation maps as shown in figure 3. The gray area inside the red boundary is the area that is not flooded. The information from the inundation map can be used as input data for the evacuation route calculation that will be described later. In addition to input for the evacuation route calculation, the inundation data could determine which hotels are not in the flood zone for evacuees or response personnel coming to the affected area to assist in recovery efforts.

**Semantic Segmentation for Flooding Detection**

The next component of our tool applies semantic segmentation, a computer vision (CV) technique which adds near real-time overhead imagery information (**McInness**, 2018). We use publicly available data from the National Oceanic and Atmospheric Administration (NOAA)'s Remote Sensing Division, which displays city and street-level images of Lumberton, NC as mosaic tile images. We used semantic segmentation to classify the images images after Hurricane Florence at the pixel level into the categories of interest such as water, infrastructure and buildings (**Ishida**, 2019 **Kahn**, 2019, **Prasad**, 2019, **Wurm**, 2019, **Van Etten**, 2018). Semantic segmentation can be used to identify a wide variety of classes or features, such as roads, trees, and buildings. In our current work we used U-Net to produce segmentation maps that identify the water / flooded segments (**Ronnenburger**, 2015) among other features. U-Nets are a specific neural network architecture that can learn from the data to identify features of interest at the pixel level by leveraging multiple feature map resolutions. In our Figure 4 example, blue regions on the segmented image represent pixels classified as water segments on the map. For the purposes of routing, we assumed that if the algorithm has detected any amount of water on a road; that road would be made inaccessible by flooding.

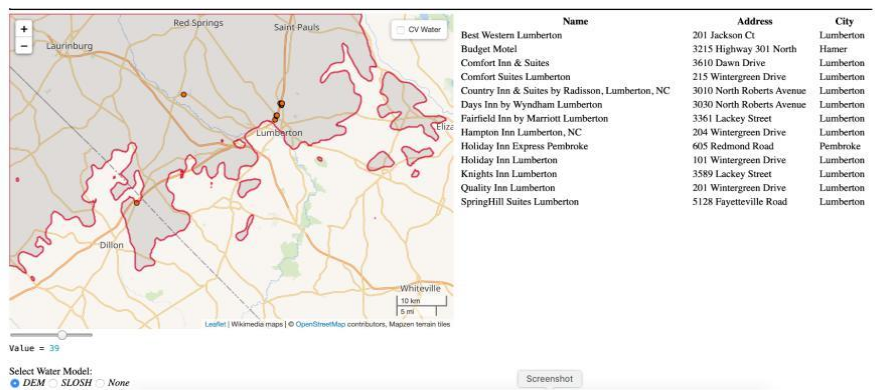

**Figure 3. (Left) Inundation map based on topography and digital elevation map. Non-shaded area is predicted to be flooded. (Right) list of available lodging options given the flooding model.**

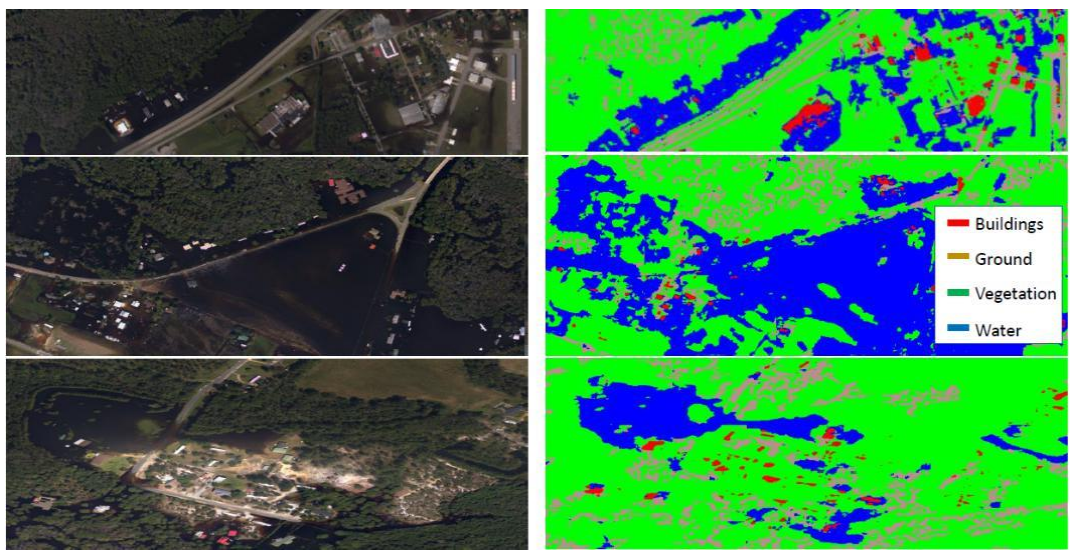

**Figure 4. Flood Mapping Before and After Hurricane Florence using Semantic Segmentation**

Whereas pixel-level classification of an image is a laborious and time-consuming task for human analysts, we quickly and inexpensively created maps of the affected area after Hurricane Florence in Lumberton, NC using semantic segmentation. For practical disaster response, CV can be applied to the maps. Using aircraft to obtain near real-time imagery and computer vision techniques produces near-real-time inundation maps. The only limitation is the availability of visible and radar images after the emergency event has happened. The images used for the analysis have such a high resolution that the speed for data ingestion and processing would be extremely laborious for human analysts. The computer vision model substantially accelerated that effort. Both the SLOSH information and the products from the image analysis become input for the route calculation algorithm. This segmentation information can be used in a manner as described above to identify hotels or evacuation shelters not lying in the flood zone (Figure 3).

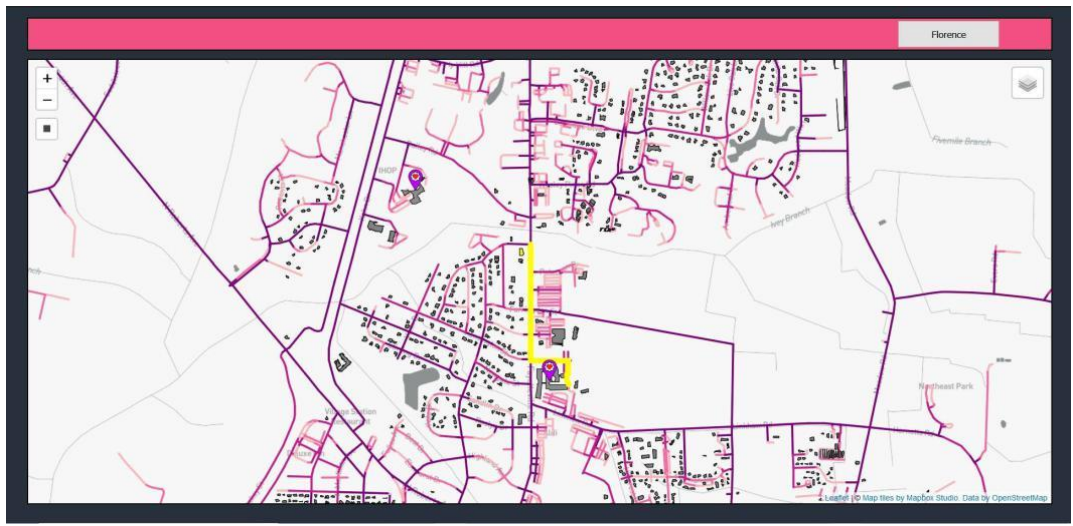

**Figure 5. Route Mapping Calculation of Lumberton, NC**

**AI-driven and User-Centered Front-End Visual Interface**

Now that there is transformed data about road flooding following Hurricane Florence from the inundation map and semantic segmentation, this data can be used as input data with a machine learning algorithm to calculate the shortest route from unsafe origination to a different safer destination. The AI augmentation tool combines user-centered design principles with transformed open street map data and road network strengths to calculate the shortest path calculations using any number of machine learning algorithms, including the Dijkstra algorithm. Building locations are necessary inputs to calculate evacuation routes. The building geometries we used for the routing component in prior research were extracted using (**Open Street Maps Overpass API**, 2019). However, certain areas – such as the affected area of Lumberton, NC - do not have well represented data in OSM. For these areas, we depend on the Google Maps page for the building data. In this situation, our approach is to search for the region of interest in Google Maps, zoom into the map at a depth of at least 17 (the minimum zoom level to observe building features), and then computationally extract the locations of the building by using the RGB values associated with the building and road features (241,241,241). The resulting geometries provide us with accurate information about the location of each building.

The novelty of that AI augmentation component is that it takes a collection of data layers, including computer vision-derived products and provides a crucial visual user interface to present the information to emergency personnel. Unlike existing GPS maps, our interface provides another more in-depth layer of information about locations (i.e. - a building is a hurricane shelter). This component highlights and communicates the safest evacuation route from a user-selected origination point to a safer end point and can be used before, during or after the natural hazard happens. This framework uses a human-centered design to drive technical requirements for a decision-maker responsible for emergency communications. A secondary functionality of the tool is to provide a list of available lodging options given the latest flooding maps produced by the computer vision model. Available lodging can then be incorporated into the routing framework as additional safer destinations.

A user needs to perform three distinct actions, or clicks in order to find the best route to safety. In this iteration of the mapping user interface, we added a gray button with the emergency name in the top right corner. First, a user clicks on the gray Florence button and actual flood data from Hurricane Florence from the inundation map and semantic segmentation is loaded on top of existing data as an additional information layer. As seen in figure 5, then a user can click on the yellow building on Peterson Drive near Peterson's Toyota, which is their origination point. Next, a user clicks on their preferred destination or shelter building, shown as Lumberton High School with a purple leaflet on the map north of the Aldi Grocery Store. After the user completes these three interactions with the map, the shortest path is calculated based on road network strength described in previous research. The yellow highlighted route is the recommended route the user should take to get to a safe destination or evacuation shelter. In this use case, a user exits their building, proceeds South on Fayetteville Road past Farringdom Street, then proceeds west on Kensington Street arriving at the evacuation shelter. If a route is closed due to flooding or other circumstances, that segment can be deleted and a backup evacuation route is calculated.

**FUTURE WORK**

Future iterations of the system will include applying design framework improvements for another User Type - the Evacuee. The Evacuee might have slightly different usability and technical requirements than emergency personnel and therefore potentially need a different design framework. Further usability studies may reveal other additional features that would be helpful for both user types such as labeling text or simple word instructions in multiple languages within the tool. We need to investigate how to update the map in near-real time during the various phases of the natural hazard. Furthermore, there are common infrastructure challenges such as lack of electricity and Internet that will affect successful implementation of the tool at scale. The next iteration of a visual tool might include additional offline features that could be used in the event of these common infrastructure challenges or other features that ingest data about an event based on a mobile device's location and suggest actions for the user.

We might want to consider additional considerations for the *evacuation*, *hazard* and *recovery* phases of an emergency that are not currently included. Future work will include additional data that addresses the hazard risk faced by populations in an emergency context, as well as including other spatial data that captures other hydro-logical features, road incline and other spatio-temporal vehicle traffic data. We also feel our work could be used for a variety of other use cases specific to other types of natural hazard or emergency situations where people or things are forced to move from one location to a different destination. There may additionally be other use cases outside of the domain of humanitarian logistics such as city planning when construction closes certain portions of roads or where users want multiple layers of information about a location in the same place on a map. Additionally, we would want to deploy this system prior to, during and after a disaster event to evaluate its effectiveness, responsiveness and the user experience with it.

**CONCLUSION**

Artificial Intelligence is being used to solve complex problems in both the corporate and private sectors. In this work we use it for social good with an emphasis on combining AI Augmentation technology with other components of planning and relief efforts in humanitarian logistics. Our emergency tool can be used in the context of managing risks in crisis and emergencies and can be used by different types of users to plan for natural hazards and communicate evacuation scenarios to improve resilience. Our tool is one example of how technology can reduce human loss risk in the context of humanitarian logistics, and also how machines can be used in times of great need to augment human decision-making capabilities. We believe there is still much more that can be done with this technology but this is an initial step to help and improve human lives.

Our tool combines inundation maps and semantic segmentation to improve information given to emergency personnel before, during and after Hurricane Florence. This information can augment human decision-making and reduce the uncertainty created in emergency contexts. Improved decision-making increases community resilience before, during and after an emergency. By combining transformed layers of heterogeneous data with hydro-logical inundation models and computer vision semantic segmentation, our system provides emergency response personnel and other stakeholders with timely access to relevant, comprehensive and reliable information. Data on road flooding and other infrastructure damage related to the emergency event is automatically loaded into the route analysis machine learning algorithm to provide a fast, near-real-time data source to inform disaster response and communications. We demonstrate the feasibility of our system after Hurricane Florence in September 2018 for the city of Lumberton, NC. Our current system can improve route planning, and allow emergency personnel to initiate contingency plans by providing near real-time infrastructure-damage assessments. We focus on safety and assisting people in already chaotic situations by emphasizing AI capabilities to learn and redirect people as terrain may change during a natural hazard.

**BIBLIOGRAPHY: REFERENCES AND CITATIONS**

Bosch, M., Christie, G. Gifford, C., 'Sensor Adaptation for Improved Semantic Segmentation of Overhead Imagery' in: Proceedings of the IEEE Winter Conference on Applications of Computer Vision, WACV, https: //arxiv.org/abs/1811.08328, 2019

Carswell, W. and Lukas, V., 'The 3D Elevation Program—Flood risk management: U.S. Geological Survey Fact Sheet', pp. 3018-3087, https://doi.org/10.3133/fs20173081, 2018

Hilljegerdes, M. and Augustijn-Beckers, E., 'Evaluating the Effects of Consecutive Hurricane Hits on Evacuation Patterns in Dominica. Published' in: Proceedings of the 16th International Conference on Information Systems for Crisis Response and Management, https://ris.utwente.nl/ws/files/120016684/1734_MartinHilljegerdes1_Ellen_WienAugustijn_Beckers22019.pdf, 2019

'Hurricane Florence', https://en.wikipedia.org/wiki/Hurricane_Florence, 2018

Ishida H., Matsutani K., Adachi M., Kobayashi S., and Miyamoto R., 'Intersection Recognition Using Results of Semantic Segmentation for Visual Navigation', 2019

Kahn, L., Ortiz-Ulloa, B., Pavon-Harr, V. Savas, O., 'Artificial Intelligence (AI) Augmentation for Humanitarian Logistics: A Framework and Visual Tool for Planning and Communicating Optimal Evacuation Routes for Natural Hazards' in: Proceedings of the AAAI Fall Symposium, 2019

Kapucu, N., Hawkins, C., and Rivera, F., 'Disaster resilience: Interdisciplinary perspectives', 2013

McInnes, Leland, John Healy, and James Melville. 'Umap: Uniform manifold approximation and projection for dimension reduction.' arXivpreprintarXiv:1802.03426, 2018

National Oceanic and Atmospheric Administration, 'The National Water Model', https://water.noaa.gov/ about/nwm, 2018

Prasad, L., Pope, P. and Sentz, K., 'Semantic segmentation of multispectral overhead imagery' https://www.semanticscholar.org/paper/Semantic-segmentation-of-multispectral-overhead-Prasad-Pope/1ba6daebbd2e43858d5af648424577ca3be559bf, 2016

Ronnenberger, O., Fischer, P. T. Brox, 'U-Net: Convolutional Networks for Biomedical Image Segmentation' https://arxiv.org/pdf/1505.04597.pdf, 2015.

US Geological Survey, 'Digital Elevation Models' https://www.usgs.gov, 2018

US Geological Survey, 'Flood Inundation Map Program' https://www.usgs.gov, 2018

US Geological Survey, 'Three Dimensional Elevation Program' https://www.usgs.gov, 2018

Van Etten, A., Lindenbaum, D. and Bacastow, T., 'SpaceNet: A Remote Sensing Dataset and Challenge Series', https://arxiv.org/abs/1807.01232, 2018

Van de Walle, B. and Turoff, M., 'Decision support for emergency situations' in: Handbook on Decision Support Systems, International Handbook on Information Systems Series, 2007

Wikipedia, 'Overpass API - Open Street Map', https://wiki.openstreetmap.org/wiki/Overpass_API, 2019

Wurm, M., Stark, T., Zhu, X., Weigandad, M. and Taubenböcka, H., 'Semantic segmentation of slums in satellite images using transfer learning on fully convolutional neural networks' https://www.sciencedirect.com/ science/article/pii/S0924271619300383, 2019

Zachry, B. C., W. J. Booth, J. R. Rhome, and T. Sharon, 'A National View of Storm Surge Risk and Inundation' in: Weather, Climate, and Society, vol: 7(2), pp. 109-111, http://dx.doi.org/10.1175/WCAS\T1\textendashD\T1\textendash14\T1\textendash00049.1, 2015